\documentclass[
	twocolumn,
	amsmath,
	amssymb,
	prx,
	aps,
    superscriptaddress, 
	nofootinbib
	]{revtex4-2}
 
\usepackage{graphicx}
\usepackage{xcolor}
\usepackage{amsmath}
\usepackage{amssymb}
\usepackage{booktabs}
\usepackage{titlesec}
\usepackage{hyperref}

\begin{document}

\title[Article Title]{Dynamic Avalanches: Rate-Controlled 
Switching and Race Conditions}

\author{Lishuai~Jin}
\email{lishuai.jin@cityu.edu.hk}  
\affiliation{AMOLF, Science Park 104, 1098 XG Amsterdam, The Netherlands}
\affiliation{Huygens-Kamerlingh Onnes Lab, Universiteit Leiden, 2300 RA Leiden, The Netherlands}
\affiliation{Department of Mechanical Engineering, City University of Hong Kong, Hong Kong, China.}

\author{Martin~van Hecke}
\affiliation{AMOLF, Science Park 104, 1098 XG Amsterdam, The Netherlands}
\affiliation{Huygens-Kamerlingh Onnes Lab, Universiteit Leiden, 2300 RA Leiden, The Netherlands}

\begin{abstract}
Avalanches are rapid cascades of rearrangements driven by cooperative flipping of hysteretic local elements. Here we show that flipping dynamics and race conditions---where multiple elements become unstable simultaneously---give rise to dynamic avalanches that cannot be captured by static models of interacting elements. We realize dynamic avalanches in metamaterials with controlled flipping
times and demonstrate how this allows us to modify, promote, and direct avalanche behavior. Our work elucidates the crucial role of internal dynamics in complex materials and introduces dynamic design principles for materializing targeted pathways and sequential functionalities.
\end{abstract}

\maketitle
Quasistatic driving of multistable media, including crumpled sheets~\cite{matan2002crumpling,shohat2022memory}, amorphous solids~\cite{reichhardt2023reversible,fiocco2014encoding}, and metamaterials~\cite{meeussen2023multistable,overvelde2015amplifying}, gives rise to dynamical orbits
marked by sequences of rapid, irreversible transitions between metastable states. Under cyclic driving, these transitions form pathways that encode sequential shape morphing~\cite{melancon2022inflatable}, memory effects~\cite{paulsen2024mechanical,paulsen2019minimal,keim2011generic,paulsen2014multiple,lindeman2021multiple,keim2021multiperiodic}, and computational capabilities of complex media  \cite{van2021profusion,kwakernaak2023counting,liu2024controlled}.
Hence, key challenges include uncovering how the underlying physics shapes these pathways, 
identifying which pathways and combinations of pathways are physical,
and realizing (meta)materials with tailored pathways for targeted functionalities \cite{ding2022sequential,kamp2024reprogrammable,jules2022delicate,kwakernaak2023counting,bense2021complex,liu2024controlled,jin2024multiperiodic,el2024tunable,sirote2024emergent, hyatt2023programming,jin2020guided,raney2016stable}.

Transitions often entail the flipping of bistable or hysteretic elements such as ridges~\cite{meeussen2023multistable}, particle clusters~\cite{wang2022propagating}, beams~\cite{liu2024controlled}, bistable reaction-diffusion cells \cite{vergassola2018mitotic},
or hysteretic springs~\cite{shohat2024geometric}, which allows one to distinguish two types of transitions.
Transitions composed of a single flipping event are referred to as elementary, and those composed of multiple events are avalanches. These events occur in rapid succession, suggesting that avalanche characteristics may depend on the driving rate, the dynamics of individual flipping events, and the time delay between flipping events due to the finite velocity of signal propagation. For example, when the driving rate becomes comparable to the internal timescales, recent studies have shown that dynamic driving protocols can significantly influence the resulting avalanches \cite{lindeman2023competition,jules2023dynamical,liu2021delayed,huang2024exploiting}. 

Here, instead, we consider asymptotically slow driving and show that the internal dynamics can lead to  
{\em dynamic avalanches} 
that cannot be described in a quasistatic framework. 
First, underdamped flipping dynamics can lead to vibrations and dynamical orbits that overshoot a local energy minimum, thus promoting the formation of avalanches. Second, when the time delay between flipping events is small, race conditions may occur in which one flipping event destabilizes multiple other elements simultaneously \cite{huffman1954synthesis,teunisse2024transition,szulc2022cooperative}; the order in which these unstable elements flip then determines the avalanche.

We experimentally realize dynamic avalanches in metamaterials 
composed of hysteretic springs in which we control the flipping dynamics, and use simulations to reveal their underlying phase-space structures \cite{liu2024controlled,shohat2024geometric}.
We contrast our results with current models of interacting hysterons \cite{van2021profusion,lindeman2021multiple,keim2021multiperiodic}. Finally, we leverage our insights to design a metamaterial that exhibits six complex avalanches, enabling a challenging pathway that mimics a three-bit binary counter.
Our work reveals the crucial role of internal timescales in slowly driven systems and paves the way for designer materials with novel pathways for advanced sensing and computation.

\begin{figure}[h]
\centering
\includegraphics[width=\columnwidth]{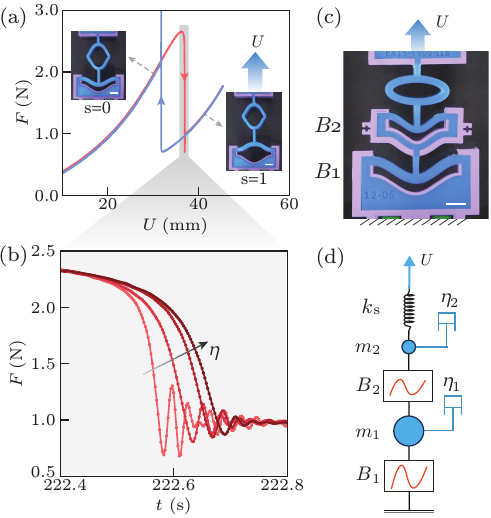}
\caption{(a) Force-extension response of a hysteretic element; insets show typical
configurations for $s=0$ and $s=1$. (b) Close-up of the $0\rightarrow 1$ transition dynamics in air (light) and in silicon oils for viscosities $\eta=350, 500, 1000$ cSt  (increasingly dark).
(c) Sample 1 is a metamaterial containing two hysteretic elements ($B_1, B_2$) coupled in series with an (elliptical) spring (scale bar 1 cm). (d) Numerical model of dynamic hysteretic elements, with spring constant $k_s$, masses $m_1$ and $m_2$, and viscous dampers $\eta_1$ and $\eta_2$.}
\label{fig1}
\end{figure}

{\em Hysteretic metamaterials.---}
Networks of hysteretic springs \cite{liu2024controlled,shohat2024geometric} are promising platforms to investigate and realize dynamic avalanches. We materialize hysteretic springs
using encased curved beams that flip between internal phase $s=0$ (relaxed U-shape), and $s=1$ (snapped shape) under tension (Fig.~1a) \cite{liu2024controlled}.
Experimentally, we control their flipping timescale through (partial) immersion in viscous fluids (Fig.~1b; see Appendix A). 
We focus on $n$ hysteretic springs coupled in series for which the static properties, interactions, design, and modeling are well understood (Fig.~1c)~\cite{liu2024controlled,shohat2024geometric}, and in parallel
perform simulations of a realistic model of dynamic hysteretic springs (Fig.~1d) \cite{lindeman2023competition,jules2023dynamical}.
Under controlled deformation $U$, serial coupling 
leads to antiferromagnetic-like interactions, capable of producing avalanches \cite{liu2024controlled,shohat2024geometric}.

{\em Dynamic Avalanches under Slow Driving.---}
We start by demonstrating that dynamic effects can promote avalanches.
We represent the collective states of our metamaterial as \( S = (s_1, s_2, \dots) \) and define the up and down switching thresholds at which element \( i \) becomes unstable when the control parameter \( U \) is increased or decreased as \( U_i^\pm(S) \).
We focus on a metamaterial made of two hysteretic springs and examine the transition triggered by destabilizing state \( (10) \) as the driving \( U \) is decreased just below the switching threshold \( U_1^-(10) \). 
In the simplest case, this results in an elementary transition \( (10) \rightarrow (00) \), which requires state \( (00) \) to be stable at \( U = U_1^-(10) \)\footnote{We tacitly assume \( U_2^+(00) < U_1^+(00) \).}, leading to the necessary condition:
\begin{equation} 
G := U_1^-(10) - U_2^+(00) < 0~,
\end{equation}
where $G$ is the gap between the relevant switching thresholds.
However, when the flipping of the first element destabilizes the second element and causes it to flip upward from \( 0 \) to \( 1 \), the avalanche $(10)\!\rightarrow(01) $ occurs. 
A sufficient condition for this avalanche is that $G>0$, and in current hysteron models, this condition is taken as both sufficient and necessary \cite{van2021profusion,teunisse2024transition,baconnier2024proliferation,mungan2019networks,mungan2019structure}.

To show that destabilization of the second element by the first can be promoted by dynamic effects, 
we design sample 1 with a small but negative value of $G$. We slowly sweep the global deformation $U$ (rate 10 mm/min), measure all transitions and their associated switching thresholds, and collect these in a transition graph (t-graph) \cite{paulsen2019minimal,regev2021topology,keim2021multiperiodic,lindeman2021multiple,van2021profusion}.
We first perform experiments when the sample is 
immersed in Polydimethylsiloxane oil (viscosity: 350 cSt) and then when the sample is suspended in air (Fig.~2; Appendix A). While the t-graph of the sample in oil is free of avalanches, the t-graph of the sample in air exhibits the avalanche  $(10) \!\rightarrow \! (01)$ despite $G<0$  \footnote{The values of the switching thresholds in air and in water are slightly different due to experimental errors.} (Fig.~2).
The observation of this avalanche and its dependence on the damping
strongly points to the role of dynamic effects.
Indeed, high-speed imaging of this avalanche reveals that the flipping dynamics of element one lead to an overshoot that triggers the switching of element two, just before element one reaches its '0' configuration (see Supplemental Material~\cite{SI}, Movie S1).

\begin{figure}[t]
\centering
\includegraphics[width=\columnwidth]{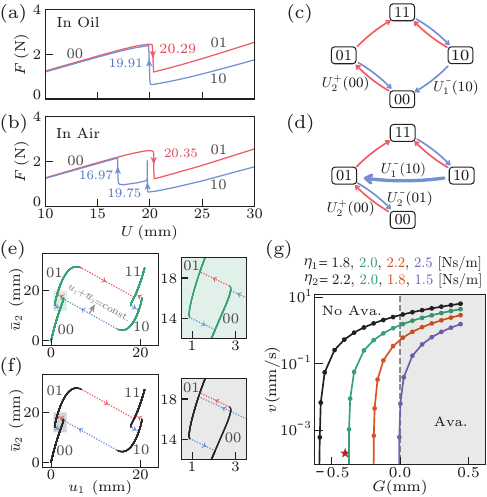}
\caption{
(a,b) Force-extension curves for sample 1 submersed in oil (a) and air (b). The switching thresholds  $U_2^+(00)$ and  $U_1^-(10)$ for the  up and down transitions of states $(00)$ and $(10)$ are indicated (as well as $U_2^-(01)$ in (b)). Both exhibit a negative $G$. Note the overshoot of the force during the $(10)\rightarrow (01)$ avalanche at $U=19.75$ mm. 
(c,d) Corresponding t-graphs, with the associated switching thresholds indicated.
(e,f) Orbits of the numerical model with  
$v:=\dot{U}=2\cdot10^{-4}$ mm/s,  $G=-0.4$ mm, $m_i=(20,10)$ g
and $\eta_i=(2.0,2.0)$ Ns/m (e) and  $\eta_i=(1.8,2.2)$ Ns/m (f), where $\bar{u}_2=u_2+u_k$, and $u_1+\bar{u}_2=U$.
Zoom ins: In (e), dashed lines represent transitions
$(10)\rightarrow(00)$, $(01)\rightarrow(00)$, and $(00)\rightarrow(01)$;
in (f), the transition $(10)\rightarrow(00)$ is replaced by $(10)\rightarrow(01)$.
(g) Avalanche vs no-avalanche parameter regime for designs with the same masses, different pairs of $\eta_i$ and range of $G$ and $v$ (star indicates $G$ and $v$ for panels (e,f)). Dynamic avalanches occur on the right side of each curve, including in regimes where the hysteron model predicts the absence of 
avalanches (white).
}\label{fig2}
\end{figure}

To further clarify how internal flipping dynamics can promote dynamic avalanches, even in the limit of quasistatic driving, we use a realistic numerical model which describes the evolution of the extensions $u_i$ of hysteretic springs coupled to masses and viscous dampers (Fig.~1d; see Appendix B and Supplemental Material, Sec. II). The spring potentials control the switching thresholds and thus $G$, and the masses $m_i$ and damping coefficients $\eta_i$ control the internal dynamics \cite{jules2023dynamical, lindeman2023competition}. 
First, we fix the driving rate and masses, design the elements to feature
a small negative gap \( G \), and compare the trajectories \( u_i(t) \) resulting from destabilizing state \( (10) \) by lowering \( U \), for two different values of \( \eta_i \) (Fig.~2e,f). For $\eta_i=(2.0,2.0)$ Ns/m, there is no avalanche and the trajectory lands on \( (00) \), where
we note that since $G$ has a small value, the trajectory lands near the edge of instability of state $(00)$ (Fig.~2e). In contrast, 
for $\eta_i=(1.8,2.2)$ Ns/m the orbit overshoots $(00)$ and instead reaches state $(01)$ (Fig.~2f). This illustrates how dynamic effects can lead to overshooting
state $(00)$, in particular when its range of attraction is small, thus triggering a dynamic avalanche (Supplemental Material, Movie
S2).

We then study how the critical value of \( G \), which separates the avalanche and no-avalanche regimes, depends on the driving rate \( v \) across a range of viscosity ratios \( \eta_i \). Our data show that as \( v \downarrow 0 \), the critical value of \( G \) asymptotes to a constant, and that avalanches can occur for \( G < 0 \) over a wide range of viscosities (Fig.~2g). These avalanches are most strongly promoted when \( \eta_1 / \eta_2 \ll 1 \), while for \( \eta_1 \gg \eta_2 \), the critical value of \( G \) approaches zero: in this regime, element two switches much faster than element one, allowing the system to quickly settle into the intermediate state. 
This shows that dynamical effects can promote, but not suppress avalanches, and that the internal flipping dynamics remain important even in the limit of quasistatic driving. Moreover, this diagram indicates that, away from the transition curve, dynamic avalanches are robust to changes in design ($G$), damping, and driving rate.

\begin{figure}[t]
\centering
\includegraphics[width=1\columnwidth]{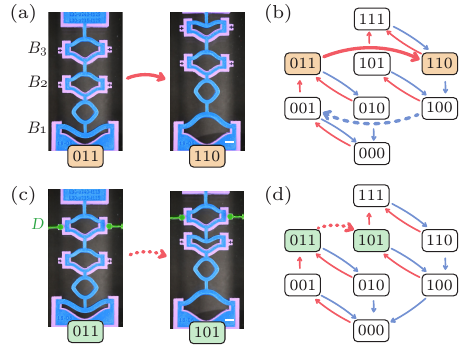}
\caption{\textbf{Race conditions.}
Sample 2 features three serially coupled hysteretic elements ($B_1-B_3$), 
a spring between
$B_1$ and $B_2$, and the third element can be damped (green).
(a-b) Avalanche transition of 
sample 2 in air, and corresponding t-graph (for switching thresholds, see Supplemental Material, Fig.~S13). The relevant gap for the $(011)\rightarrow(110)$ avalanche is positive, whereas the gap for $(100)\rightarrow(001)$ avalanche is negative, so that the latter is a dynamic avalanche which violates condition {\em(i)}  
i.e., the intermediate state 000 is stable (dashed).
(c-d) Avalanche transition of sample 2, where element three is damped by submersing rigidly attached wings ($D$, green) in oil, and the corresponding t-graph. The $(011)\rightarrow(101)$ dynamics avalanche violates condition (ii) as it does not follow the scaffold (dotted). Scale bars, 1 cm.
}\label{fig3}
\end{figure}

{\em Race Conditions and Dynamic Steering.---} 
We now consider race conditions, where one flipping event destabilizes multiple other elements simultaneously \cite{huffman1954synthesis,teunisse2024transition,szulc2022cooperative}\footnote{Race conditions are distinct from degeneracies that occur when multiple switching thresholds are equal.}. 
In random systems, 
the likelihood of race conditions increases with 
the number of elements
\cite{baconnier2024proliferation}. 
In designed systems, such race conditions can be realized straightforwardly.
The simplest race condition requires three elements, and we investigate the case where flipping element one of state $(011)$ destabilizes both elements two and three, and where both states $(110)$ and $ (101)$ are stable at $U=U_1^+(011)$. Dynamic effects—particularly localized damping—then can determine which state will be reached  (for the corresponding energy landscape, see Supplemental Material, Fig.~S9).

This race condition is materialized in sample 2 by making element one significantly larger than elements two and three: 
when, starting from state $(011)$, element one snaps from 0 to 1, this significantly compresses and destabilizes both other elements.
We verified this by selectively blocking either element two or three and observing that the other element became unstable (see Supplemental Material, Fig.~S8).

Removing the blockades, we show that 
the resulting avalanche can be selected by localized viscous damping. In air, the sample 2 exhibits the
$(011)\rightarrow(110)$ avalanche (Fig. 3a-b), whereas 
when the third element is submersed in oil, sample 2 
exhibits the $(011)\rightarrow(101)$ avalanche transition (Fig. 3c-d). 
High-speed imaging shows that the flipping of element one leads to compression of element two and then three, where the delay originates from the finite propagation speed of the perturbation induced by the first snap. In air, even though element two is transiently compressed, eventually element three flips, thus releasing the pressure on element two, whereas in oil, the damping of element three slows down its evolution and forces element two to flip (Supplemental Material, Movie S3).
Numerical simulations of this race condition scenario are consistent with the experimental observations and show that both pathways robustly occur over a range of viscosities and designs. Moreover, we find that moving the location of springs and masses additionally tunes the transitions, stressing the importance of the propagation of information through the sample (Supplemental Material).
Together, race conditions allow dynamic effects to qualitatively steer avalanche pathways, and allow the realization of t-graphs that are inconsistent with all current hysteron models (Fig.~3d).

{\em Quasistatic and Sequential Models.---}
We now compare the observed avalanches to those of current hysteron models, which model the elements as strictly binary elements \cite{van2021profusion,keim2021multiperiodic,teunisse2024transition}. 
Their interactions are encoded via the state dependence of the 
switching thresholds $U_i^\pm(S)$, which sometimes  
can be modeled theoretically   \cite{keim2021multiperiodic,lindeman2021multiple,lindeman2025generalizing}. 
In the absence of internal dynamical rules, the switching thresholds 
fully determine the transitions and avalanches. For 
race conditions, which become prevalent in large systems \cite{baconnier2024proliferation},  the model is sometimes deemed ill-defined \cite{van2021profusion} or a phenomenological rule---like flip the most unstable hysteron---is employed \cite{keim2021multiperiodic,baconnier2024proliferation}.
Avalanches then proceed by sequential flipping of hysterons, thus visiting intermediate states that are separated by one flipping event.
This sequential assumption has two consequences.
{\em (i)} Avalanches only proceed when these intermediate states are unstable, such that condition $G>0$ is both necessary and sufficient.
{\em (ii)} Avalanches and transitions follow the same scaffold \cite{teunisse2024transition}; in other words, 
if the system exhibits the elementary transition $S_1\rightarrow S_2$, then an avalanche where $S_1$ is an intermediate state also evolves to $S_2$. 

Clearly, sample 1 and sample 2 in air violate condition {\em (i)}, while the partially damped sample 2 violates condition {\em (ii)} (Fig.~3d). To see this, note that the up avalanche starting from $(011)$ 
has the intermediate state $(111)$, which itself has an elementary transition to $(110)$. Hence, the avalanche $(011)\rightarrow (110)$ is consistent with the scaffold (Fig.~3b), but the 
avalanche $(011)\rightarrow (101)$ displayed by the partially immersed sample 2 is not (Fig.~3d).  Hence, qualitative and quantitative features of dynamic avalanches are inconsistent with sequential, quasistatic models.

\begin{figure}[t]%
\centering
\includegraphics[width=1.0\columnwidth]{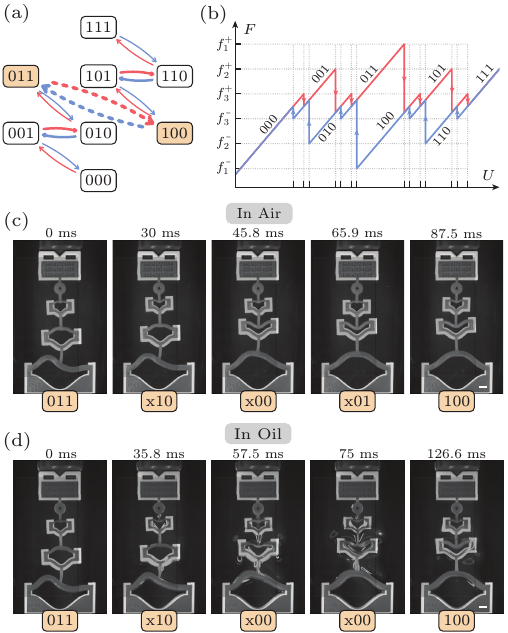}
\caption{\textbf{Three-bit counter.} (a) T-graph of a three-bit counter. (b) Target curve $F(U)$. (c,d) High-speed images during the avalanche transition in air and in silicone oil. Scale bars, 1 cm.
}\label{fig5}
\end{figure}

{\em Three-bit counter.---}
We next harness race conditions and  dynamic effects to create a metamaterial that exhibits a challenging pathway inspired by a three-bit counter, i.e., evolves as $(000)\leftrightarrow (001) \leftrightarrow (010) \leftrightarrow \dots \leftrightarrow (111)$, which
requires six avalanches
(Fig.~4a).
Crucially, the pair of avalanches 
$(011)\leftrightarrow(100)$
require race conditions; as shown previously, sequential avalanches for serially coupled hysteretic elements are restricted to those where two elements change phase in the opposite direction
\cite{liu2024controlled}.
To realize this pathway, we define a target curve for the collective force-displacement $F(U)$ (Fig.~4b), and consider the design problem of serially coupled bilinear elements
each described as $f_i=k_i u_i - g_i s_i$, where $g_i$ are the force jumps resulting from the switching of the $i$-th hysteron (Appendix, Fig. A1b) \cite{liu2024controlled,shohat2024geometric}.
This target curve can be realized with three serially coupled elements with 
hierarchically organized 'sizes', i.e., where the 
switching thresholds satisfy $f_1^-<f_2^-<f_3^-<f_3^+<f_2^+<f_1^+$, where $g_1>g_2>g_3$,
and where race conditions allow the challenging avalanches 
$(011)\leftrightarrow(100)$
(see Appendix C).

We realize sample 3 based on such considerations,
using three elements of increasing size (Fig.~4c-d). Its $F(U)$ curve satisfies the design target, including the nested ordering of the switching forces $F^\pm(S)$ and the correct sequential ordering of all switching thresholds $U^\pm(S)$. 
Although the detailed evolution of the sample during the anomalous avalanche $(011)\leftrightarrow(100)$ depends on whether the sample is in air or in oil, in both cases, the elements switch simultaneously thus evidencing the race condition.
This demonstrates that race conditions can be leveraged to realize (meta)materials with challenging pathways (Fig.~4a and Supplemental Material, Movie S4).

{\em Discussion and Outlook.---}
Although the underlying static energy landscape determines the states, their stability, and the initial mode of evolution after an instability, the physical dynamics are crucial to fully capture the irreversible transitions between states \cite{lindeman2023competition,jules2023dynamical}. 
In general, there are three overall time scales at play - the inverse global strain rate $\dot{\gamma}^{-1}$, the 
typical time for an avalanche to cross one element $\tau^w$, and
the switching times of the elements $\tau^s_i$, where in particular the latter can vary with each element. When $\dot{\gamma}^{-1} \gg \tau^w L \gg \tau^s_i$, where $L$ is the linear size in number of elements, and when the dynamics are overdamped,
all time scales are separated, race conditions are irrelevant, and flipping events occur in isolation, with each event forming a smooth connection  between intermediate states
--- although we note that even in this limit, the sequence of flipping may be determined by the spatial separation between elements, something that is missing in current hysteron models. But when 
the condition $\tau^w L \gg \tau^s_i$ is no longer satisfied, or when the flipping events are underdamped, the trajectories between states are intrinsically dynamical, with
controlled flipping dynamics and race conditions allowing one to materialize avalanches that quantitatively and qualitatively deviate from sequential hysteron models.

Traveling waves in linear arrays of identical bistable elements with nearest neighbor interactions
—observed in both biological systems \cite{vergassola2018mitotic} and mechanical metamaterials \cite{jin2020guided,raney2016stable}—are closely related to the dynamic avalanches discussed here. We propose that underdamped dynamics may promote such waves, just as they promote avalanches. However, since race conditions require interactions beyond nearest neighbors, our second mechanism may not be relevant for traveling waves.

Our dynamic design strategy imbues metamaterials with unprecedented sensing capabilities, where avalanches depend on 
ambient viscosities (Fig.~2), or where more complex situations, such as race conditions, could be tuned to compare different viscosities in materia (Fig.~3). Moreover, at fixed damping, our dynamic strategy allows controlled avalanches for accelerated shape-morphing \cite{jin2020guided}, autonomous soft robots \cite{kamp2024reprogrammable},
and designer pathways (that are inconsistent with current hysteron models), thus aiding the design of metamaterials with
emergent memory and computing \cite{liu2024controlled, melancon2022inflatable,paulsen2024mechanicalhys}.

\begin{acknowledgments}
{We thank D. Shohat, P. Baconnier, J. Liu, and M. Liu
for insightful discussions, and D. Ursem for technical support.  L.~J. and M.~v.~H. acknowledge funding from the European Research Council Grant ERC-101019474.
}
\end{acknowledgments}

\bibliography{refs}

\section*{Appendix}
\setcounter{figure}{0}
\renewcommand{\thefigure}{A\arabic{figure}}

\appendix


\section{Experimental Methods and Samples}

We realize multistable metamaterials by coupling $n$
bistable, hysteretic springs in series with an elliptical-shaped, nearly-linear spring, which facilitate sharp snapping transitions \cite{liu2024controlled}. 
Each hysteretic spring consists of a pre-curved flexible beam 
encapsulated within a rigid 3D-printed clamp. The flexible beams of the samples are fabricated from a two-component siloxane rubber (Mold Star 30, Smooth On) using standard molding techniques, while the rigid components consist of 3D-printed polylactic acid (PLA) frames produced with an Ultimaker S3 printer. 

Isolated elements switch up ($s_i:0\rightarrow1$) 
when their extension $u_i$ exceeds the bare switching threshold $u_i^+$,
and switch down ($s_i:1\rightarrow0$) when $u_i$ falls below
$u_i^-$. The beams' thickness $t_i$, span $L_i$ and the amplitude of the beam's profile $A_i$ determine these switching thresholds. 
For detailed design parameters for each of our samples, see Supplemental Material.

The mechanical response of our samples is characterized using a horizontal tensile testing machine (Instron 5965), which provides precise control over axial displacement U with an accuracy better than 4 $\mu$m. A 100 N load cell is employed to measure the applied force with a resolution down to 0.5 mN. All experiments are conducted at a constant displacement rate of 10 mm/min. During testing, the samples are imaged using both a Canon camera (EOS 800D) and a high-speed camera (Phantom v4.2) to capture their 
dynamic evolution.
The samples are tested either in air or fully or partially immersed in Polydimethylsiloxane oil (viscosity: 350 cSt) for various damping conditions. To implement partial submersion of a hysteron, we mount two wings (green parts in Fig. 3c) to its sides. The wings are then immersed in silicone oil while the rest of the sample remains clear of the oil, creating a partially damped environment for the sample.

\section{Numerical Model}
To investigate the dynamical behavior of coupled bistable elements,
we perform simulations of a simple mechanical model introduced recently
\cite{lindeman2023competition,jules2023dynamical}.
This model represents networks of bistable springs, linear springs, masses and viscous dampers
(Fig.~1d). First, each bistable spring is represented with a cubic force-displacement curve $f_i(u_i)$. Second, the elliptical spring in our samples is represented by a linear spring $f_k(u_k)$. The detailed forms of $f_i(u_i)$ and $f_k(u_k)$ can
be obtained from experimental data of our samples, or can be adjusted to represent the range of stability of the states \cite{lindeman2023competition,jules2023dynamical,liu2024controlled,shohat2024geometric}.  Together, networks of these elements describe the stable and unstable collective states and exhibit hysteretic transitions between stable states. 

To model how the transition pathways depend on 
dynamical effects, the model is augmented with masses and viscous dampers. For example, sample 1, consisting of two hysteretic  elements coupled to an elliptical ring, is modeled as follows:
\begin{align}\label{eq_goven}
    \nonumber
    &m_1 \ddot{x_1}+\eta_1 \dot{x_1}+f_1(u_1)-f_2(u_2)=0,\\
    \nonumber
    &m_2 \ddot{x_2}+\eta_2 \dot{x_2}+f_2(u_2)-f_k(u_k)=0,\\
    &u_1+u_2+u_k=U,
\end{align}
where $m_i$ represents the masses that are coupled to viscous dampers characterized by $\eta_i$ (Fig.~1d). For details, see Supplemental Material.

\section{Design of Three-Bit Counter}
\begin{figure}[t]
\centering
\includegraphics[width=1.\columnwidth] {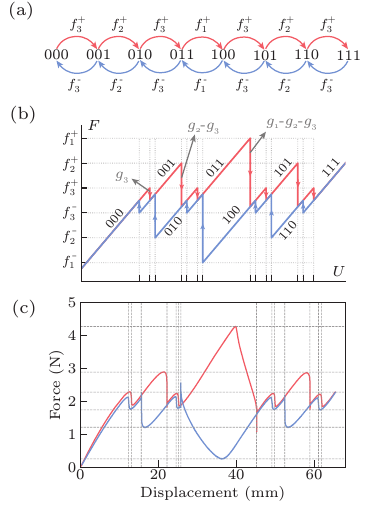}
\caption{\textbf{Design of three-bit counter.} (a) The transitions between subsequent states must occur at one of the six critical switching forces. (b) Target curve $F(U)$. (c) Experimentally observed $F(U)$ curve for sample 3, with critical forces and displacements highlighted. 
}\label{fig5}
\end{figure}

We consider the design of a targeted pathway that materializes a three-bit binary counter (Fig.~\ref{fig5}a). We start from
a simple model of
three serially coupled bilinear elements, each described by \cite{liu2024controlled,shohat2024geometric}:
\begin{equation}
f_i=k_i  u_i - g_i  s_i~.
\end{equation}
Here $f_i$ and $u_i$ are the forces and extensions, $k_i$ is the stiffness, and $g_i$ is the force jump between $s_i=0$ and $s_i=1$ branches; each element switches between these branches when
$u_i$ exceeds the switching thresholds $u_i^\pm$, or equivalently, when the force exceeds the critical forces $f_i^\pm$.
Mechanical balance equations dictate that the global force $F=f_1=f_2=f_3$, and that the global extension $U=u_1+u_2+u_3$, which, under controlled extension $U$, allows to construct the $F(U)$ curve \cite{liu2024controlled,shohat2024geometric}.

The target $F(U)$ curve is constructed such that, for each value of \( U \), at most two stable states exist  (Fig.~\ref{fig5}b). This ensures unambiguous transitions that do not rely on the dynamic effects of type (i) to be controlled.
Moreover, the target pathway specifies the sequence of flipping events, and as each flipping event is associated with a specific critical force $f_i^\pm$, this implies that these have to be nested as in Fig.~\ref{fig5}b, i.e.
\begin{equation}
\label{eq:nested}
f_1^-<f_2^-<f_3^-<f_3^+<f_2^+<f_1^+~.
\end{equation}
Similarly, the 14 critical values of $U$ satisfy a chain of inequalities of the form
$U^-_3(001)<U^+_3(000)< \cdots<U^-_3(111)<U^+_3(110)$,
and translating these to the switching forces and force drops, and using elementary algebra, yields a second set of design equations (where we assume $k_i=1$ without loss of generality)
\begin{eqnarray} \label{eq:chain}
& f_3^- \!+\!g_3 < f_3^+ < f_2^- \!+\!g_2 < f_2^+ \!+\!g_3 & \dots \nonumber 
 \\  \dots < &f_3^- \!+\!g_2\!+\!g_3 <f_3^+ \!+\! g_2 <f_1^- \!+\! g_1 & \dots \nonumber \\ 
 \dots <  &f_1^+ \!+\!g_2\!+\!g_3< f_3^- \!+\!g_1 \!+\!g_3 <f_3^+ \!+\!g_1& \dots \nonumber \\ 
 \dots < &f_2^- \!+\!g_1\!+\!g_2 < f_2^+ \!+\!g_1\!+\!g_3& \dots \nonumber \\
 \dots < &f_3^- \!+\!g_1\!+\!g_2\!+\!g_3< f_3^+ \!+\!g_1\!+\!g_2~.& \\ \nonumber
\end{eqnarray}

There is a range of choices for the switching forces and force drops $g_i$---which both can be designed experimentally at the level of individual elements---that satisfy these equations. 
For example, an equidistant ordering of the switching fields requires rapidly growing values of the 
force drop $g_i$ with the significance of the elements.
Hence, in our experimental design we take each element to satisfy the nested constraints on the critical forces, and the rapidly increasing values of the force drops. After some iterative design, this produced sample 3, which satisfies both sets of design equations (Fig.~\ref{fig5}c). Note that while sample 3
exhibits sharp transitions between states in the correct order, 
the deformation of the largest element exhibits higher beam modes, which modify $f_1(u_1)$ and make the collective curve $F(U)$ deviate from the strictly piecewise linear target - we stress that the sequence of transitions is correct. This 
deviation shows the robustness of our design strategy to such experimental deviations.

\end{document}